\documentclass[12pt,preprint]{aastex}

\shorttitle{planet formation in turbulent disks} 

\shortauthors{Ida et al.}

\begin{document}

\title{Accretion and destruction of planetesimals in turbulent disks}

\author{Shigeru Ida}
\affil{Tokyo Institute of Technology,
Ookayama, Meguro-ku, Tokyo 152-8551, Japan}
\email{ida@geo.titech.ac.jp}

\and 

\author{Tristan Guillot, Alessandro Morbidelli}
\affil{Observatoire de la C\^{o}te d'Azur,
CNRS UMR 6202, BP 4229, 06304 Nice Cedex 4, France}
\email{guillot@oca.eu, morby@oca.eu}

\begin{abstract}
We study the conditions for collisions between planetesimals to be
accretional or disruptive in turbulent disks, through analytical
arguments based on fluid dynamical simulations and orbital
integrations.  In turbulent disks, the velocity dispersion of
planetesimals is pumped up by random gravitational perturbations from
density fluctuations of the disk gas.  When the velocity dispersion is
larger than the planetesimals' surface escape velocity, collisions
between planetesimals do not result in accretion, and may even lead to
their destruction. In disks with a surface density equal to that of
the ``minimum mass solar nebula'' and with nominal MRI turbulence, we
find that accretion proceeds only for planetesimals with sizes above
$\sim 300$\,km at 1AU and $\sim 1000$\,km at 5AU.  We find that
accretion is facilitated in disks with smaller masses. However, at
5AU and for nominal turbulence strength, 
km-sized planetesimals are in a highly erosive regime
even for a disk mass as small as a 
fraction of the mass of Jupiter. The existence of
giant planets implies that either turbulence
was weaker than calculated by standard MRI models 
or some mechanism was capable of producing Ceres-mass planetesimals 
in very short timescales. 
In any case, our results show that in the presence of
turbulence planetesimal accretion is most difficult in massive
disks and at large orbital distances.
\end{abstract}

\keywords{solar system: formation --- planets and satellites: formation 
    --- accretion, accretion disks --- turbulence}

\section{Introduction}

It is often considered that the evolution of protoplanetary disks and
the consequent accretion of gas by the central protostar are driven by
turbulent viscosity due to a Magneto-Rotational-Instability (MRI)
\citep[e.g.,][]{bal91}.  \citet{lau04} and \citet{nel04} carried out
fluid dynamical simulations of MRI and found that the random torques
due to the turbulent density fluctuations give rise to a random walk
in semimajor axes of planetesimals.  \citet{ric03} pointed out through
model calculations that the random walk expands the effective feeding
zone of protoplanets, and may lead to rapid formation of large cores
for gas giants.  Through a Fokker-Planck treatment, \citet{johnson06}
also pointed out the importance of the random walk in planet accretion.  
Adopting the semi-analytical formula for the random torque 
derived by \citet{lau04}, \citet{ogi07} performed
N-body simulations for the late stages of terrestrial planet
accretion with a disk significantly depleted in gas, starting from
Mars-mass protoplanets.
They found that the MRI
turbulence indeed helps to reduce the number of
accreted terrestrial planets which is otherwise too large compared to
our Solar System.

However, \citet{nel05} found through direct integrations of
the orbits of protoplanets in a MRI turbulent disk that orbital
eccentricities are also excited.
\citet{britsch08} also found a similar feature 
in a self-gravitating disk.
While the random walk itself is favorable to
the growth of protoplanets by avoiding isolation, the
excitation of their eccentricities, which had been neglected in
\citet{ric03} and \citet{johnson06}, 
is a threat for planetesimal accretion processes
because of increased collision velocity. 
Unfortunately, \citet{nel05}'s orbital integrations were 
limited to 100--150 Keplerian times
and neglected collision processes, so that it is not possible to
conclude from that work whether planetesimals should grow or be eroded
in the presence of turbulence. 

In the present article, we explore by which paths planetesimals may
have grown to planet-sized bodies in turbulent disks. 
Because the level of density fluctuations due to the MRI turbulence is
not well determined, we choose to study the qualitative effects
of the turbulence on the accretion of planetesimals and 
their dependence on the key parameters of the problem, in particular
the progressive removal of the gas disk. 
In \S 2, we summarize the conditions for the accretion and destruction
of planetesimals in terms of their orbital eccentricities. 
In \S 3, we analytically derive the equilibrium eccentricity
for which the excitation due to turbulence is balanced
by damping due to tidal interactions with the disk gas,
aerodynamic gas drag, and collisions.
Comparing the equilibrium eccentricities with critical 
eccentricities for accretion and destruction,
we derive critical physical radii and masses of
planetesimals for accretion or destruction.
The results are applied to viscously evolving disks (\S 4).
We then discuss possible solutions for the problem of the formation of
planetesimals and planets (\S 5).

\section{Accretion and destruction conditions}

We summarize the accretion and destruction conditions below.  From
energy conservation, the collision velocity ($v_{\rm coll}$) between
two planetesimals (labeled 1 and 2) satisfies
\begin{equation}
E = \frac{1}{2}v_{\rm coll}^2 - \frac{G(M_1 + M_2)}{R_1 + R_2}
= \frac{1}{2}v_{\rm rel}^2,
\end{equation}
where $M_j$ and $R_j$ are the mass and physical radius of a
planetesimal $j$ ($j=1,2$), and $v_{\rm rel}$ is their relative
velocity when they are apart from each other.  When the velocity
dispersion of planetesimals $v_{\rm disp}$ is larger than the Hill
velocity that is given by $(M/3M_{\ast})^{1/3} v_{\rm K}$, where
$M_{\ast}$ is the mass of the host star and $v_{\rm K}$ is
the Keplerian velocity, $v_{\rm rel}$ is
approximated by $v_{\rm disp}$ \citep[e.g.,][]{IN89,I90}.  The total
energy then becomes:
\begin{equation}
2E = v_{\rm coll}^2 - v_{\rm esc}^2 \simeq v_{\rm disp}^2,
\end{equation}
where $v_{\rm esc}$ is the (two-body) surface escape velocity
defined by
\begin{equation}
v_{\rm esc} = \sqrt{ \frac{2G(M_1 + M_2)}{R_1 + R_2} }.
\end{equation}

Collisional dissipation decreases the energy by some fraction of
$v_{\rm coll}^2/2$.
If $v_{\rm disp} \ll v_{\rm esc}$, the 
collisional dissipation results in $E < 0$ after collision.
On the other hand, $E$ is likely to be still positive 
after a collision with $v_{\rm disp} \gg v_{\rm esc}$.
Thus, for moderate dissipation,
the condition for an accretional collision is
$v_{\rm disp} < v_{\rm esc}$ \citep[e.g.,][]{ohtsuki93}.
Since the orbital eccentricity $e \simeq v_{\rm disp}/v_{\rm K}$, 
a collision should result in accretion for
$e < e_{\rm acc}$, where
\begin{equation}
\begin{array}{lll}
e_{\rm acc} & \simeq \frac{v_{\rm esc}}{v_{\rm K}}
 & \simeq 0.28 \left(\frac{M}{M_{\oplus}}\right)^{1/3}
  \left(\frac{\rho_{\rm p}}{3{\rm gcm}^{-3}}\right)^{1/6}
  \left(\frac{r}{1{\rm AU}}\right)^{1/2} \\
 & & 
\simeq 0.036 
  \left(\frac{R}{10^3{\rm km}}\right)
  \left(\frac{\rho_{\rm p}}{3{\rm gcm}^{-3}}\right)^{1/2}
  \left(\frac{r}{1{\rm AU}}\right)^{1/2}.
\end{array}
  \label{eq:e_acc}
\end{equation}
In the above relation, $\rho_{\rm p}$ is the bulk density of
the planetesimals and
$M \sim M_j$ (for simplicity, $M_1 \sim M_2$ is assumed).
The physical radius $R$ is given by
\begin{equation}
R=7.8 \times 10^8 (M/M_{\oplus})^{1/3} (\rho_{\rm p}/3{\rm gcm}^{-3})^{-1/3}
{\rm cm}.
\end{equation} 

A collision results in destruction if the collision velocity
is such that the
specific kinetic energy of a collision ($v_{\rm col}^2/2$) exceeds
\begin{equation}
Q_{\rm D} \simeq 
  \left[ Q_0 
  \left(\frac{R}{1{\rm cm}}\right)^{a} + 
  3 B \left(\frac{\rho_{\rm p}}{3{\rm gcm}^{-3}}\right) 
  \left(\frac{R}{1{\rm cm}}\right)^{b} \right] {\rm erg/g},
\label{eq:Q_D}
\end{equation}
where $Q_0$ is the material strength, 
$B \simeq 0.3$--2.1, $a \simeq -0.4$, and $b \simeq 1.3$ 
\citep{Benz_Asphaug99}.
For basalt rocks or water ice, 
$Q_0 \simeq 10^7$--$10^8$ \citep{Benz_Asphaug99}, but it can take
a significantly smaller value for loose aggregates.
\footnote{For porous materials, $Q_0$ is rather higher (W. Benz,
private communication).}

We adopt $Q_0 = 3 \times 10^7$ as a nominal value.
Self-gravity (the second term in the r.h.s.) dominates the material
strength when $R \ga 100$ m. 
In this regime, adopting $B \sim 1$, 
a collision results in destruction for
$e > e_{\rm dis}$, where 
\begin{equation}
\begin{array}{ll}
e_{\rm dis} \simeq \frac{\sqrt{2 Q_{\rm D}}}{v_{\rm K}}
 & \simeq 0.50 \left(\frac{M}{M_{\oplus}}\right)^{0.22}
  \left(\frac{\rho_{\rm p}}{3{\rm gcm}^{-3}}\right)^{0.28}
  \left(\frac{r}{1{\rm AU}}\right)^{1/2} \\
 & \simeq 0.13 \left(\frac{R}{10^3{\rm km}}\right)^{0.65}
  \left(\frac{\rho_{\rm p}}{3{\rm gcm}^{-3}}\right)^{0.5}
  \left(\frac{r}{1{\rm AU}}\right)^{1/2}. 
\end{array}
\label{eq:e_D}
\end{equation}

\section{Equilibrium eccentricities}

We first derive the equilibrium eccentricities of
planetesimals at which the excitation by the MRI turbulence is
balanced by damping due to drag and/or collisions.  
Comparing the
estimated eccentricities with $e_{\rm acc}$ and $e_{\rm dis}$, 
we then evaluate the outcome of
collisions between planetesimals as a function of planetesimal size,
turbulent strength, and surface density of disk gas.

For an easy interpretation, we provide in this section analytical
relations based on a model in which the gas and
solid components of disk surface density are scaled with the
multiplicative factors $f_g$ and $f_d$:
\begin{equation}
\Sigma_{g} = 2400 f_g \left(\frac{r}{1\mathrm{AU}}\right)^{-3/2} \,\mathrm{g\, cm}^{-2},
\label{eq:Sigma_g}
\end{equation}
and 
\begin{equation}
\Sigma_{d} = 10 f_d \eta_{\rm ice} \left(\frac{r}{1\mathrm{AU}}\right)^{-3/2} \,\mathrm{g\, cm}^{-2},
\label{eq:Sigma_d}
\end{equation}
where $\eta_{\rm ice} \simeq 3$--4
is an enhancement factor of $\Sigma_d$ due to ice condensation.
If $f_g=f_d=1$, $\Sigma_{g}$ and $\Sigma_d$ are 
1.4 times those of the minimum mass solar nebula model \citep{hayashi81}.

In this section, we also use the disk temperature distribution 
obtained in the optically thin limit \citep{hayashi81},
\begin{equation}
T \simeq 280
\left(\frac{r}{1\, {\rm AU}}\right)^{-1/2}
\left(\frac{L_*}{L_{\odot}} \right)^{1/4}
\; {\rm K},
\label{eq:T_opt_thin_disk}
\end{equation}
where $L_*$ and $L_{\odot}$ are the stellar and solar luminosities,
respectively. The corresponding sound velocity is
\begin{equation}
c_s = 1.1 \times 10^5 \left( \frac{r}{1\,{\rm AU}} \right)^{-1/4}
\left(\frac{L_*}{L_{\odot}} \right)^{1/8}
\; {\rm cm/s}.
\label{eq:sound_velocity_hayashi}
\end{equation}
Since the disk scale height is given by $h = \sqrt{2}c_s/\Omega_{\rm K}$
(assuming that $T$ is vertically uniform in the disk),
eqs.~(\ref{eq:Sigma_g}) and (\ref{eq:sound_velocity_hayashi})
yield the disk gas density at the midplane as
\begin{equation}
\rho_g = \frac{\Sigma_g}{\sqrt{\pi}h}
       = 2 \times 10^{-9} f_g
         \left( \frac{r}{1\, {\rm AU}} \right)^{-11/4}
         \; {\rm gcm}^{-3}.
\label{eq:rho_midplane}
\end{equation}

\subsection{Excitation}

The orbital eccentricities of planetesimals are pumped up both by the
random gravitational perturbations from density fluctuations of disk
gas, as well as by mutual gravitational scattering among
planetesimals.  Assuming planetesimals have equal masses, their
orbital eccentricities should be excited to at most $\sim e_{\rm acc}$
by the mutual scattering \citep[e.g.,][]{Safronov69}.  
As will be shown below, the value of
this excentricity is smaller than that due to the turbulent
excitation, except for very large planetesimals ($\sim 10^3$\,km or
larger), and/or in the case of significantly depleted gas disks.
For simplicity, in this work we choose to neglect the possibility that mutual
scattering dominates over turbulent excitation. It is therefore important to
note that our results may be slightly optimistic when concerning the
possibility of accretion of massive planetesimals. 

\begin{figure}
\epsscale{0.8}
\plotone{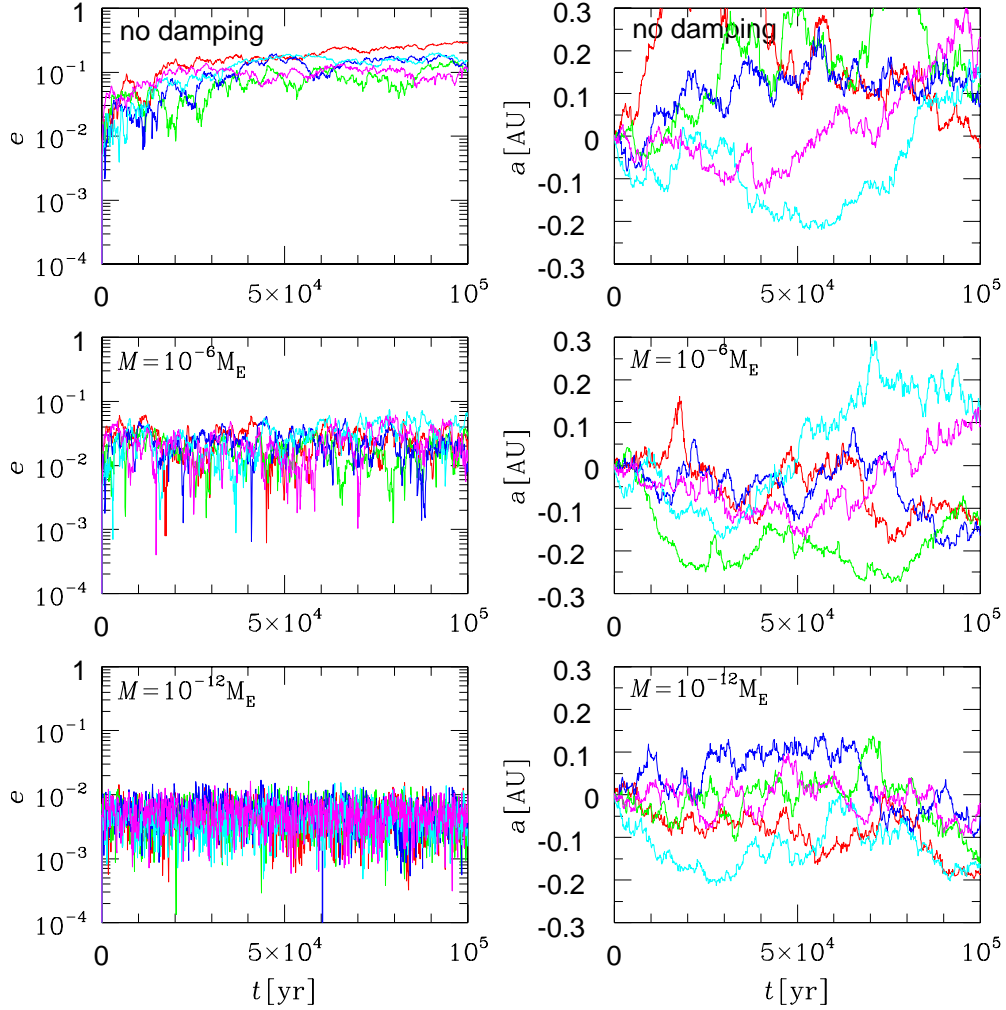}
\caption{Evolution of the eccentricity (left panels) and of the
  semimajor axis (right panels) as a function of time, 
  for different planetesimal masses 
  (top to bottom). 
  A single planetesimal is integrated in a turbulent disk and
  5 independent runs 
  with different random number seeds 
  for the generation of turbulent density fluctuations are shown in each panel.
  {\it Top panels\/}:
  We consider no tidal and gas drag. 
  {\it Middle panels\/}: Results
  including gas drag for planetesimals of
  $M=10^{-6}M_{\oplus}$. 
  {\it Bottom panels\/}: Results
  for planetesimals of $M=10^{-12}M_{\oplus}$. In all cases, we
  assume an initial orbital distance of 1AU, $f_g=1$ and
  $\gamma=0.01$.}
\label{fig:e_evol}
\end{figure}

The orbital eccentricities that result from the turbulent density
fluctuations in the disk are provided by \citet{ogi07} on the basis of
orbital integrations with empirical formula by \citet{lau04}, as
\begin{equation}
e \sim 0.1 \gamma 
  \left(\frac{\Sigma_g}{\Sigma_{g,1}}\right)
  \left(\frac{r}{1{\rm AU}}\right)^{2}
  \left(\frac{t}{T_{\rm K}}\right)^{1/2}
  =
  0.1 f_g \gamma 
  \left(\frac{r}{1{\rm AU}}\right)^{-1/4}
  \left(\frac{t}{1{\rm year}}\right)^{1/2},
\label{eq:e_random}
\end{equation}
where $\Sigma_{g,1}$ is $\Sigma_{g}$ at 1AU with $f_g = 1$
(eq.~[\ref{eq:Sigma_g}]) and
$\gamma$ is a non-dimensional parameter to express
the disk turbulence.
\footnote{Although \citet{ogi07} suggested that eq.~(\ref{eq:e_random})
  may be enhanced by a factor 10 by the inclusion of $m=1$ modes, the $m=1$
  modes actually enhance only the amplitude of random walk in
  semimajor axis ($\Delta a$) but not the eccentricity.  
  Since higher $m$ modes fluctuate over shorter timescales, 
  they tend to cancel out on the orbital period of a planetesimal.
  For these modes, $\Delta a/a$, which is due to time variation 
  of the potential, is much smaller than $\Delta e$, because the 
  latter is also excited by the non-axisymmetric structure.
  The inclusion of slowly varying $m=1$ modes enhances 
  $\Delta a/a$ up to the order of $\sim e$.
  On the other hand, the definition of $\Gamma$ in eqs.~(5) and
  (34) of \citet{ogi07} should be multiplied by $\pi$.  We use
  eq.~(\ref{eq:e_random}) for the eccentricity excitation, which is
  consistent with an orbital calculation including $m=1$ modes
  (figure~\ref{fig:e_evol}).}
Although \citet{ogi07} showed the results only at $\sim 1$AU,
we here added a dependence on $r$ using scaling arguments
(see the Appendix).
Orbital integration for other $r$ show a consistent dependence.
Note that $\Delta a/a \sim e$, where
$\Delta a$ is the amplitude of random walk in semimajor axis.  Since
$e \ll 1$, the radial distance $r$ and semimajor axis $a$ are
identified here.

From the simulation results by \citet{lau04}, the value of $\gamma$ may
be $\sim (1/3) (\delta \rho/\rho) \sim 10^{-3}$--$10^{-2}$ for MRI
turbulence.  
In this paper, we use $\gamma = 10^{-3}$ as a fiducial value.
Interestingly, with a quite
different approach, \citet{johnson06} derived a similar formula
for $\Delta a/a$ with the same dependences on $r$, $\Sigma_g$ and $t$.
If $e \simeq \Delta a/a$, their formula is consistent with ours.
They suggested that $\gamma \sim \alpha$ 
or $\alpha^{1/2} h/a$ where $h$ is disk scale height
and $\alpha$ is the parameter for the alpha prescription
for turbulent viscosity \citep{alpha}.
For $\alpha = 10^{-3}$--$10^{-2}$, 
their estimate is also similar to our fiducial value.

The top panels in fig.~\ref{fig:e_evol}
show the results of an orbital integration 
with 4-th order Hermite scheme for
the evolution of $e$ and $\Delta a/a$ 
with turbulent perturbations but without any damping.
Five independent runs with different random number seeds 
for the generation of turbulent density fluctuations 
\citep{ogi07} are plotted in each panel. 
The initial $e$ and $i$ are $10^{-6}$.
For $f_g = 1$, $\gamma = 0.01$, and $r=1$AU, as used 
in fig.~\ref{fig:e_evol}, 
eq.~(\ref{eq:e_random}) is reduced to
$e \sim 10^{-3}(t/1{\rm year})^{1/2}$.
To highlight the effect of turbulence, we used
a larger value of $\gamma$ than the fiducial value.
The evolution of the root mean squares of the five runs 
in fig.~\ref{fig:e_evol} agrees with eq.~(\ref{eq:e_random})
within a factor of $\sim 2$.
From eq.~(\ref{eq:e_random}), the excitation timescale is
\begin{equation}
\tau_{\rm exc} = \frac{e}{de/dt} 
  \simeq 2 \times 10^{2} 
  \gamma^{-2} e^2
  \left(\frac{\Sigma_g}{\Sigma_{g,1}}\right)^{-2}
  \left(\frac{r}{1{\rm AU}}\right)^{-4} T_{\rm K} 
  = 2 \times 10^{2} f_g^{-2} \gamma^{-2} e^2
  \left(\frac{r}{1{\rm AU}}\right)^{1/2} \; {\rm years}.
\label{eq:t_exc}
\end{equation}

\subsection{Damping}

The eccentricity damping processes are
i) tidal interaction with disk gas, ii) aerodynamical gas drag,
and iii) inelastic collisions.
The tidal damping timescale (i) is derived by \citet{tan04} as
\begin{equation}
\tau_{\rm tidal} \simeq
            1.3
            \left(\frac{M}{M_{\odot}}\right)^{-1}
            \left(\frac{\Sigma_{g} r^2}{M_{\odot}}\right)^{-1}
            \left(\frac{c_s}{v_{\rm K}}\right)^{4}
            \Omega_{\rm K}^{-1}
            \simeq 
 3 \times 10^{2} f_g^{-1} 
  \left(\frac{M}{M_{\oplus}}\right)^{-1}
  \left(\frac{r}{1{\rm AU}}\right)^2 \; {\rm years}.
\label{eq:t_tidal}
\end{equation}
The gas drag damping timescale (ii) is derived by \citet{Adachi76} as
\begin{equation}
\tau_{\rm drag} \simeq
 \frac{M v_{\rm disp}}{\pi R^2 \rho_g v_{\rm disp}^2}
\simeq
 2 \times 10^{4} f_g^{-1} e^{-1} 
  \left(\frac{M}{M_{\oplus}}\right)^{1/3}
  \left(\frac{\rho_{\rm p}}{3{\rm gcm}^{-3}}\right)^{2/3}
  \left(\frac{r}{1{\rm AU}}\right)^{13/4} \; {\rm years}.
\label{eq:t_drag}
\end{equation}

For simplicity, we evaluate the damping timescale due to inelastic
collision as the mean collision time of planetesimals, assuming that
all the planetesimals have the same mass $M$.  Since in the size
distribution caused by collision cascade, collisions with
comparable-sized bodies and those with smaller ones 
contribute similarly, the neglection of the size
distribution may not be too problematic.
Since we look for the conditions in
which collisions are non-accretional, we consider the case with
$v_{\rm disp} > v_{\rm esc}$.  Assuming that the gravitational
focusing factor $[1 + (v_{\rm esc}/v_{\rm disp})^2] \sim 1$, the
collision damping timescale is
\begin{equation}
\tau_{\rm coll} \simeq
 \frac{1}{n \pi R^2 v_{\rm disp}}
\simeq
 2 \times 10^{7} f_d^{-1} \eta_{\rm ice}^{-1} 
  \left(\frac{M}{M_{\oplus}}\right)^{1/3}
  \left(\frac{\rho_{\rm p}}{3{\rm gcm}^{-3}}\right)^{2/3}
  \left(\frac{r}{1{\rm AU}}\right)^{3} {\rm years},
\label{eq:t_coll}
\end{equation}
where $n$ is the spatial number density of planetesimals.
Note that $n \sim (\Sigma_d/M)/(v_{\rm disp}/ \Omega_{\rm K})$.

\subsection{Equilibrium eccentricity}

We now equate eq.~(\ref{eq:t_exc}) with 
eqs.~(\ref{eq:t_tidal}), (\ref{eq:t_drag}), and (\ref{eq:t_coll}),
respectively,
to obtain an equilibrium eccentricity for each damping process.
For simplicity and to a good approximation, the actual equilibrium
eccentricity can be approximated as the minimum of the three
equilibrium eccentricities. 
From eqs.~(\ref{eq:t_exc}) and (\ref{eq:t_tidal}), 
\begin{equation}
\begin{array}{ll}
e_{\rm tidal} & \simeq
1.2 f_g^{1/2} \gamma  
  \left(\frac{M}{M_{\oplus}}\right)^{-1/2}
  \left(\frac{r}{1{\rm AU}}\right)^{3/4} \\
 & \simeq
24 f_g^{1/2} \gamma  
  \left(\frac{R}{10^3{\rm km}}\right)^{-3/2}
  \left(\frac{\rho_{\rm p}}{3{\rm gcm}^{-3}}\right)^{-1/2}
  \left(\frac{r}{1{\rm AU}}\right)^{3/4}.
\end{array}
\label{eq:e_tidal}
\end{equation}
With eq.~(\ref{eq:t_drag}), 
\begin{equation}
\begin{array}{ll}
e_{\rm drag} & \simeq
4.6 f_g^{1/3} \gamma^{2/3}  
  \left(\frac{M}{M_{\oplus}}\right)^{1/9}
  \left(\frac{\rho_{\rm p}}{3{\rm gcm}^{-3}}\right)^{2/9}
  \left(\frac{r}{1{\rm AU}}\right)^{11/12} \\
 & \simeq
0.23 f_g^{1/3} \gamma^{2/3}  
  \left(\frac{R}{1{\rm km}}\right)^{1/3}
  \left(\frac{\rho_{\rm p}}{3{\rm gcm}^{-3}}\right)^{1/3}
  \left(\frac{r}{1{\rm AU}}\right)^{11/12}.
\end{array}
\label{eq:e_drag}
\end{equation}
For $f_g = 1$, $\gamma = 0.01$, and $r=1$AU,
eq.~(\ref{eq:e_drag}) predicts that
$e_{\rm drag} \simeq 0.045$ for $M/M_{\oplus}=10^{-6}$
and $e_{\rm drag} \simeq 0.01$ for $M/M_{\oplus}=10^{-12}$.
An orbital integration in the middle and bottom panels in 
fig.~\ref{fig:e_evol} shows that the results agree with the 
analytical estimate within a factor $\sim 1.5$.  
With eq.~(\ref{eq:t_coll}), 
\begin{equation}
\begin{array}{ll}
e_{\rm coll} & \simeq
3.2 \times 10^2 f_g (f_d \eta_{\rm ice})^{-1/2} \gamma
  \left(\frac{M}{M_{\oplus}}\right)^{1/6}
  \left(\frac{\rho_{\rm p}}{3{\rm gcm}^{-3}}\right)^{1/3}
  \left(\frac{r}{1{\rm AU}}\right)^{5/4} \\
 & \simeq
3.6 f_g (f_d \eta_{\rm ice})^{-1/2} \gamma
  \left(\frac{R}{1{\rm km}}\right)^{1/2}
  \left(\frac{\rho_{\rm p}}{3{\rm gcm}^{-3}}\right)^{5/6}
  \left(\frac{r}{1{\rm AU}}\right)^{5/4}.
\end{array}
\label{eq:e_coll}
\end{equation}

\begin{figure}
\epsscale{0.8}
\plotone{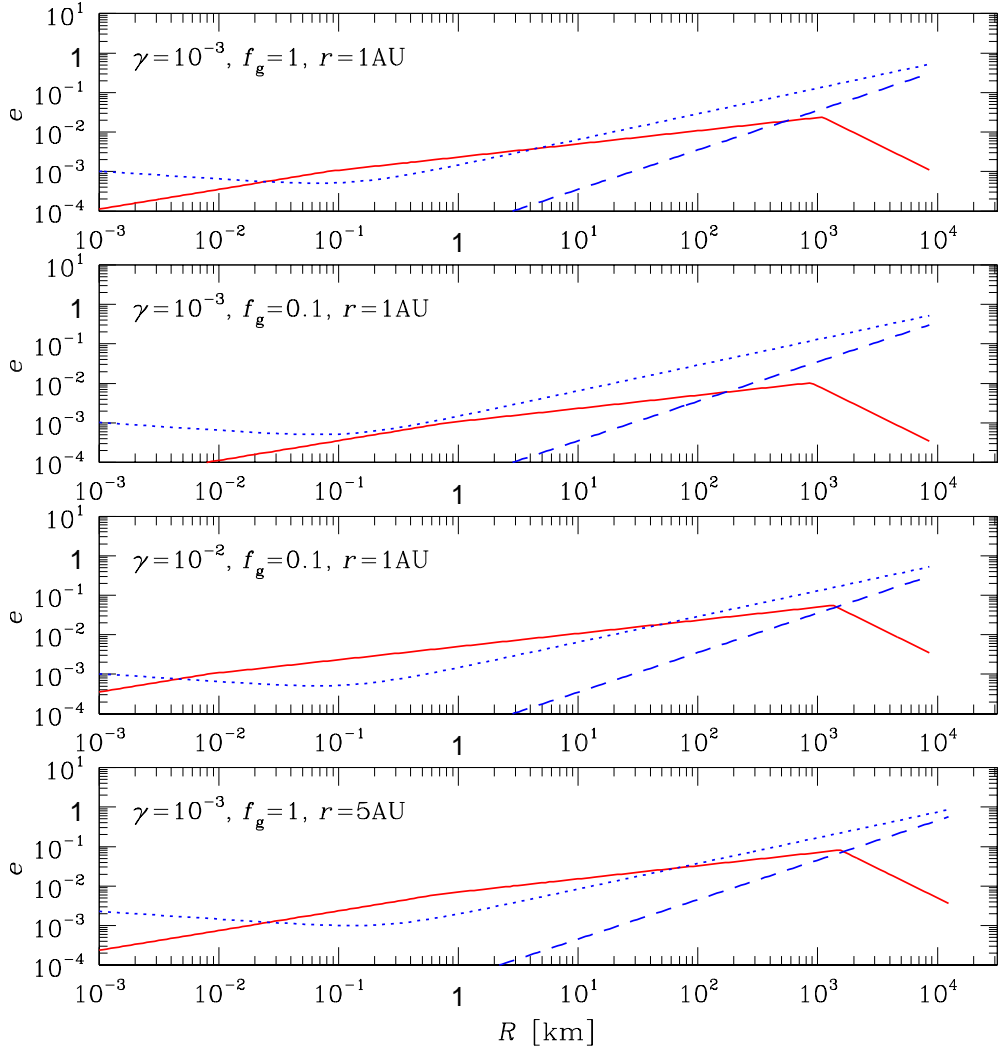}
\caption{Equilibrium eccentricities ($e_{\rm eq}$)
as a function of the physical radius $R$ of planetesimals
(solid lines).
The values of $e_{\rm eq}$ is determined by the minimum of
individual equilibrium eccentricities, $e_{\rm tidal}, e_{\rm drag}$, 
and $e_{\rm coll}$.
The critical values for accretion and destruction,
$e_{\rm acc}$ and $e_{\rm dis}$, are also plotted by
dashed and dotted lines, respectively.
At $r=1$AU, the bulk density $\rho_{\rm p} = 3 {\rm gcm}^{-3}$ is assumed,
while $\rho_{\rm p} = 1 {\rm gcm}^{-3}$ at 5 AU.
The mass of the planetesimals is given by
$M = 2.1 \times 10^{-3} (R/10^3{\rm km})^3 
(\rho_{\rm p}/3 {\rm gcm}^{-3}) M_{\oplus}$.
}
\label{fig:e_eq}
\end{figure}

In figure \ref{fig:e_eq}, the equilibrium eccentricity, $e_{\rm eq} =
{\rm min}(e_{\rm tidal}, e_{\rm drag}, e_{\rm coll})$, is plotted 
with solid lines as a function of the planetesimal radius $R$,
the corresponding planetesimal mass being $M = 2.1 \times 10^{-3}
(R/10^3{\rm km})^3 (\rho_{\rm p}/3 {\rm gcm}^{-3}) M_{\oplus}$,
Note again that the effect of mutual planetesimal
scattering is neglected.  For bodies with more than Lunar to Mars
masses, tidal damping is dominant.  This yields a
decrease in the equilibrium eccentricity with
increasing planetesimal radius for $R \ga 100$km.
For smaller mass bodies, gas drag damping dominates tidal damping
and the equilibrium eccentricity increases with increasing $R$.
For the smallest planetesimal sizes (the regions with the slightly steeper
positive gradient), collision damping is dominant, but with a
significant contribution of gas drag damping.

The limiting mass and radius at which $e_{\rm drag}$
(eq.~[\ref{eq:e_drag}]) and $e_{\rm acc}$ (eq.~[\ref{eq:e_acc}]) cross
are
\begin{equation}
\begin{array}{l}
M_{\rm acc} \simeq 
3.0 \times 10^{-4} f_g^{3/2} 
  \left(\frac{\gamma}{10^{-3}}\right)^{3}
  \left(\frac{\rho_{\rm p}}{3{\rm gcm}^{-3}}\right)^{-1/4}
  \left(\frac{r}{1{\rm AU}}\right)^{9/8} M_{\oplus}, 
\label{eq:m_acc}
\\
R_{\rm acc} \simeq 
5.2 \times 10^2 f_g^{1/2} 
  \left(\frac{\gamma}{10^{-3}}\right)
  \left(\frac{\rho_{\rm p}}{3{\rm gcm}^{-3}}\right)^{-5/12}
  \left(\frac{r}{1{\rm AU}}\right)^{3/8} {\rm km}.
\label{eq:r_acc}
\end{array}
\end{equation}
The accretion of planetesimals is possible for 
$M > M_{\rm acc}$ ($R > R_{\rm acc}$).
In the top panel of fig.~\ref{fig:e_eq}
($\gamma = 10^{-3}$ and $f_g = 1$),
planetesimal accretion proceeds in a range of
$R$'s in which the solid line ($e_{\rm eq}$) is 
located below the dashed line ($e_{\rm acc}$), that is,
only if a body is larger than Ceres.
Such large planetesimals can be formed 
by a different mechanism than pairwise accretion
such as self-gravitational instability 
in turbulent eddies \citep[e.g.,][]{Johansen07}.
When the disk gas is removed, accretion becomes
possible for smaller planetesimals 
(the 2nd panel of fig.~\ref{fig:e_eq}).
On the other hand, if turbulence is stronger
($\gamma \sim 10^{-2}$), planetesimal accretion requires
more than 1000 km-sized bodies.
This appears to be an insurmountable
barrier to accretion, even for depleted gaseous disks ($f_g \sim
0.1$), as shown in the the 3rd panel of fig.~\ref{fig:e_eq}.
Finally, at large orbital radii, planetesimal accretion is 
even more difficult (the bottom panel).

Another critical mass (radius) is the point
at which $e_{\rm drag}$ (eq.~[\ref{eq:e_drag}]) and 
$e_{\rm dis}$ in the gravity regime (eq.~[\ref{eq:e_D}]) cross,
\begin{equation}
\begin{array}{l}
M_{\rm dis} \simeq 
  6 \times 10^{-11} f_g^{3.3} 
  \left(\frac{\gamma}{10^{-3}}\right)^{6.7}
  \left(\frac{\rho_{\rm p}}{3{\rm gcm}^{-3}}\right)^{-0.6}
  \left(\frac{r}{1{\rm AU}}\right)^{4.2} M_{\oplus},
\label{eq:m_D}
\\
R_{\rm dis} \simeq 
3 f_g^{1.1} 
  \left(\frac{\gamma}{10^{-3}}\right)^{2.2}
  \left(\frac{\rho_{\rm p}}{3{\rm gcm}^{-3}}\right)^{-0.53}
  \left(\frac{r}{1{\rm AU}}\right)^{1.4} {\rm km}.
\label{eq:r_D}
\end{array}
\end{equation}
Planetesimals with $M < M_{\rm dis}$ ($R < R_{\rm dis}$) 
are disrupted by collisions
down to the sizes for which material strength is dominant
(see below).
For $\gamma \sim 10^{-3}$ and $f_g \sim 1$
(the top panel of fig.~\ref{fig:e_eq}),
planetesimals with sizes larger than several km radius survive
but without growing, while smaller planetesimals 
are disrupted.

When a planetesimal is smaller than $\sim 100$ m in size,
it is bounded by material strength rather than 
self-gravity.  
In the regime of material strength, 
$Q_{\rm D} \sim Q_0 (R/1{\rm cm})^{-0.4}$.
The body is not disrupted if
$\sqrt{2Q_{\rm D}}/v_{\rm K} > e_{\rm drag}$, which
is equivalent to
\begin{equation}
\begin{array}{l}
M \la M_{\rm mat} \simeq 0.8 \times 10^{-17} f_g^{-1.9} 
  \left(\frac{\gamma}{10^{-3}}\right)^{-3.7}
  \left(\frac{\rho_{\rm p}}{3{\rm gcm}^{-3}}\right)^{-0.9}
  \left(\frac{r}{1{\rm AU}}\right)^{2.8} M_{\oplus},
\label{eq:m_D2} \\
R \la R_{\rm mat} \simeq 16 f_g^{-0.62} 
  \left(\frac{\gamma}{10^{-3}}\right)^{-1.25}
  \left(\frac{\rho_{\rm p}}{3{\rm gcm}^{-3}}\right)^{-0.62}
  \left(\frac{r}{1{\rm AU}}\right)^{2.8} {\rm m}.
\label{eq:r_D2}
\end{array}
\end{equation}
Since in this regime, collision damping is slightly stronger than
gas drag, actual values of $M_{\rm mat}$ and $R_{\rm mat}$ are determined by
$\sqrt{2Q_{\rm D}}/v_{\rm K} > e_{\rm coll}$, so
they are slightly larger than the above estimate (see fig.~\ref{fig:e_eq}).
When $\tau_{\rm drag} \Omega_{\rm K} \la 1$, 
the planetesimals' motions are coupled
to that of the gas.  The collision velocity then cannot be expressed
in terms of orbital eccentricity. This limiting size is however
much smaller than $R_{\rm mat}$.
The collision cascade would hence stop at 
$M \sim M_{\rm mat}$ ($R \sim R_{\rm mat})$.
Regions for which the dotted lines ($e_{\rm dis}$) in
fig.~\ref{fig:e_eq} have negative gradients correspond to the material strength regime.
In the depleted disk case, $e_{\rm dis}$ is always larger than
$e_{\rm eq}$, so that the disruptive regions do not exist
(see the 2nd panel of fig.~\ref{fig:e_eq}).
Note that $R_{\rm mat} \propto Q_0^{0.94}$.
If the planetesimals are loose aggregates so that
$Q_0 < 3 \times 10^7$ (the value
for basalt rocks or water ice), the limiting size $R_{\rm mat}$ is smaller.

\section{Accretion/destruction of planetesimals in an evolving disk}

We now put these various critical physical radii in the context of the
evolution of the protoplanetary disk.  In order to investigate the
effect of departures from power-law relations of the surface density
and temperature profiles in real disks, we also present in this section
results obtained from a 1D disk model that includes an
$\alpha$-viscosity and photoevaporation \citep[see][]{Guillot06,Hueso05}. 
The parameters used in the model presented
here are a turbulent viscosity $\alpha=0.01$ and an evaporation
parameter $T_{\rm atm}=100\,$K (the temperature of the
evaporation part of the outer disk).  
Another choice of the parameters would affect
the results only marginally.

In the numerical calculation in this section, 
we evaluate the equilibrium eccentricities $e_{\rm eq}$ as a
function of planetesimal radius by solving the following relation:
\begin{equation}
\tau_{\rm exc}^{-1}=\tau_{\rm tidal}^{-1}
  +\tau_{\rm drag}^{-1}+\tau_{\rm coll}^{-1},
\end{equation}
where the different timescales are given by eqs.~(\ref{eq:t_exc}) to
(\ref{eq:t_coll}). 
The survival physical radius for accretion $R_{\rm acc}$ is then found, 
for each orbital radius in the protoplanetary
disk and for each timestep, by solving the equation,
\begin{equation}
e_{\rm eq}(R_{\rm acc})=e_{\rm acc}(R_{\rm acc}),
\end{equation}
where $e_{\rm acc}$ is given by eq.~(\ref{eq:e_acc}).
When the mean kinetic energy
is larger than the strength of a planetesimal, the 
a collision is highly erosive.
We then obtain the range $R < R_{\rm dis}$ corresponding to
the highly erosive collisions by solving the equation,
\begin{equation}
e_{\rm eq}(R_{\rm dis}) = {\sqrt{2Q_{\rm D}(R_{\rm dis})}\over v_{\rm K}},
\end{equation}
where $Q_{\rm D}$ is given by eq.~(\ref{eq:Q_D}).

\begin{figure}
\epsscale{0.7}
\plotone{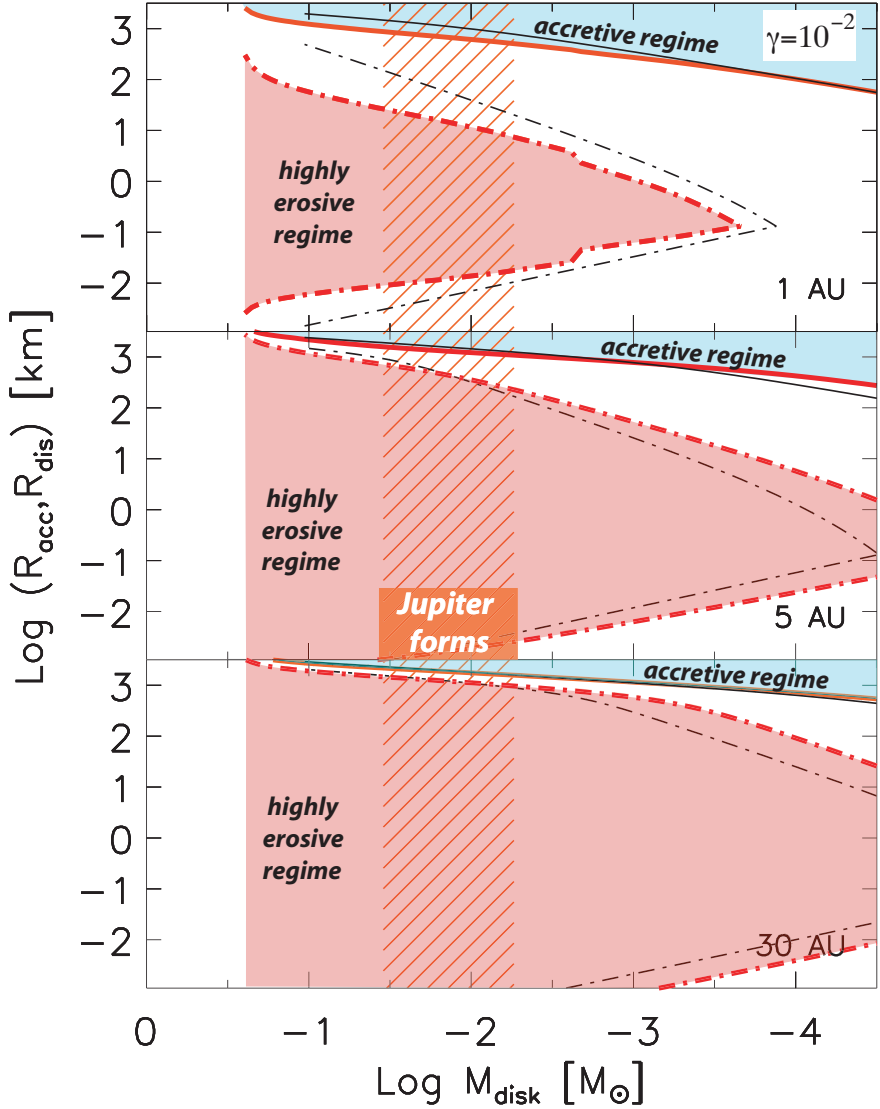}
\caption{
  Evolution of characteristic physical radii of planetesimals 
  as a function of disk mass,
  at several orbital distances in the disk: 1AU (top), 5AU (middle)
  and 30AU (bottom). 
  The solid and dot-dashed curves correspond to the boundary radius 
  for accretion regime ($R_{\rm acc}$) and to that of the highly erosive 
  regime ($R_{\rm dis}$).
  Two disk models have been used: a
  simple power-law model with $d\ln \Sigma_g/d\ln r = -3/2$
  (eq.~[\ref{eq:Sigma_g}])
  with an outer cut-off radius of 1000AU
  (thin lines), and an alpha-disk
  model with $\alpha=0.01$ and $T_{\rm atm}=100\,$K 
  \citep[see][]{Guillot06} (thick lines). The hashed region corresponds to 
  the range of disk mass (equivalently, the range of time if
  disk evolution is given)
  during which Jupiter must start accreting hydrogen/helium gas 
  (assuming it grabs between 10\% and 70\% of the disk mass 
  flux at its orbital
  distance). In the simulations, MRI turbulence is supposed to be
  high, with $\gamma=10^{-2}$. We also choose $Q_0=3\times 10^7$
  and $B=1$ (see eq.~[\ref{eq:Q_D}]). }
\label{fig:accrete_gam1d-2}
\end{figure}

\begin{figure}
\epsscale{0.7}
\plotone{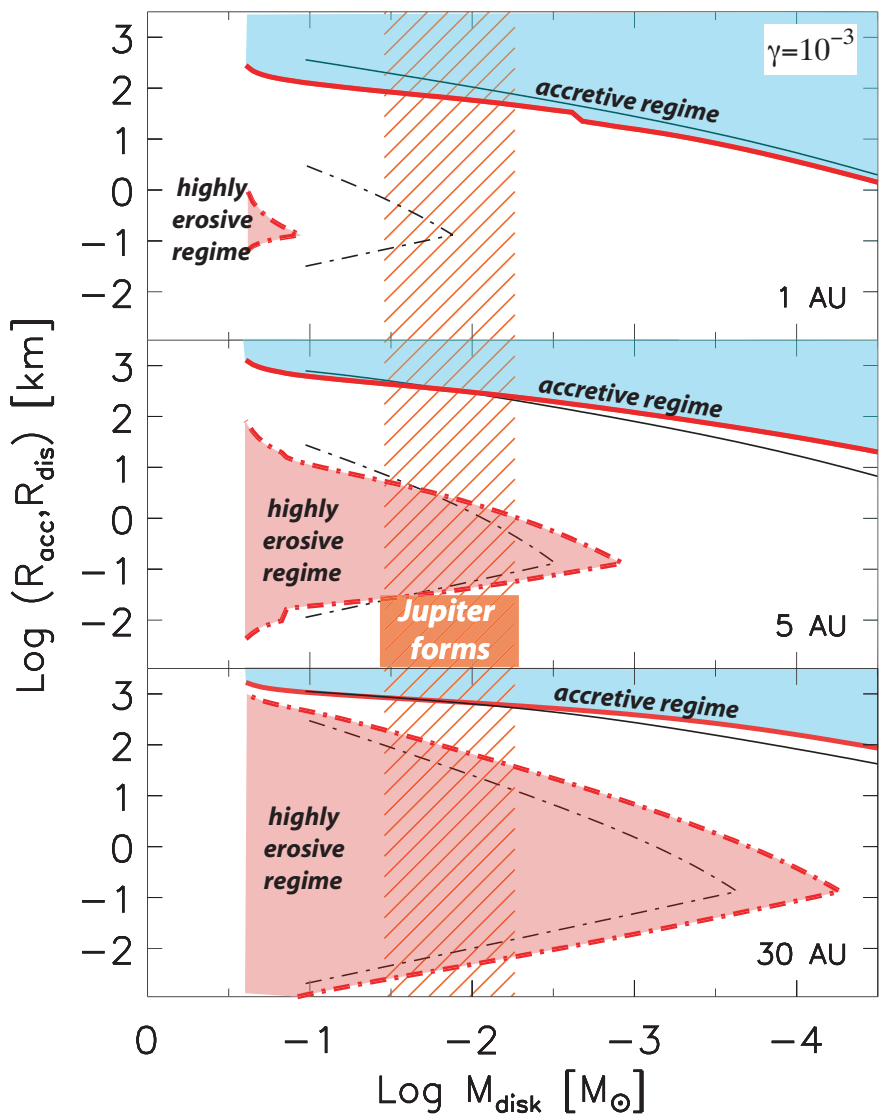}
\caption{Evolution of characteristic physical radii 
  $R_{\rm acc}$ and $R_{\rm dis}$ 
  of planetesimals as a function of disk
  mass. The parameters and labels are the same as 
  those in fig.~\ref{fig:accrete_gam1d-2} but for a medium
  turbulence case ($\gamma=10^{-3}$) (our fiducial case).}
\label{fig:accrete}
\end{figure}

\begin{figure}
\epsscale{0.7}
\plotone{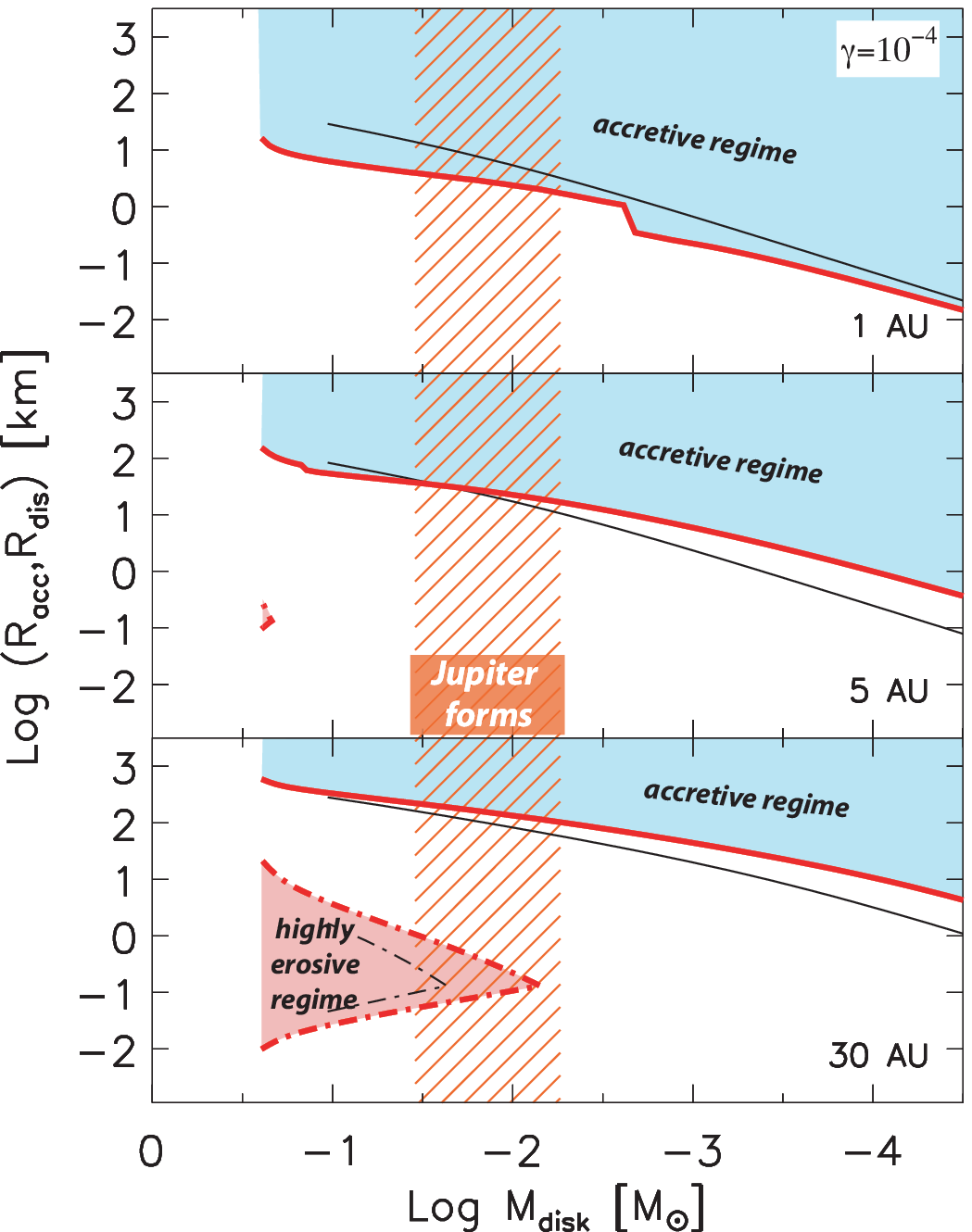}
\caption{Evolution of characteristic physical radii 
  $R_{\rm acc}$ and $R_{\rm dis}$ 
  of planetesimals as a function of disk
  mass. The parameters and labels are the same as
  those in fig.~\ref{fig:accrete_gam1d-2} but for a weak
  turbulence case ($\gamma=10^{-4}$).}
\label{fig:accrete_gam1d-4}
\end{figure}

Figures~\ref{fig:accrete_gam1d-2} to \ref{fig:accrete_gam1d-4} show
our results for three values of the turbulent excitation parameter,
$\gamma=10^{-2}$, $10^{-3}$ (our fiducial value), and $10^{-4}$. Each
figure shows, for three orbital distances, 1, 5 and 30\,AU, the
planetesimal physical radii corresponding to 
accretive and erosive regions are
plotted as a function of the total mass remaining in the disk. 
Since the disk mass decreases with time as a result of viscous 
evolution and photoevaporation, a decrease in disk mass
corresponds to evolution in time.
As shown in the previous section, planetesimal accretion
becomes easier as the disk becomes less massive simply because the
turbulent excitation, directly proportional to the
local surface density of the gas, becomes weaker. 
However, after some point, the disk becomes too light to provide a
sufficient amount of gas to form Jupiter-mass gas giants.

The figures also show as thin black lines the values obtained for a
disk that follows the slope in surface density versus orbital distance
defined for the MMSN (eq.~\ref{eq:Sigma_g})
as a function of a disk mass.
The disk mass in this model is given by 
$3.4 \times 10^{-2} f_g (r_{\rm edge}/100{\rm AU})^{1/2} M_{\odot}$,
where $r_{\rm edge}$ is the outer edge radius of the disk.
Although the original MMSN model by \citet{hayashi81} used 
$r_{\rm edge} = 35$ AU,
we here adopt $r_{\rm edge} = 1000$ AU
(for comparison, our fiducial alpha disk model with $T_{\rm atm}=100$\,K
extends up to a maximum of 350\,AU). 

These are found to be in excellent agreement with the analytical 
expressions
derived in the previous section, with small differences arising from
the simplifications inherent to the analytical approach. 
Larger differences are found between the power-law disk and the
$\alpha$-disk models mostly because of the difference in slopes
($d\ln\Sigma/d\ln r=-3/2$ for the former, $\sim -1$ for the
$\alpha$-disk) which implies that a given disk mass does not
correspond to the same surface density with two models, 
the difference being larger
at smaller orbital distances. 
However, the qualitative features of accretive and erosive
regions are similar to each other.
It should be noted that the models also
differ in their temperature profiles, but this is found to be less
important.

In the highly turbulent regime presented in
fig.~\ref{fig:accrete_gam1d-2}, a self-sustained regime of accretion
becomes possible only when planetesimals have become very
large/massive, with sizes generally well over 100km. This case also
yields a sustained area of high erosion where
the average kinetic energies of
planetesimals are above their internal energies. It is
difficult to imagine how planetary cores can form in this context
especially if they have to grow large enough to form giant planets. 

With smaller perturbations from the turbulent disk
(fig.~\ref{fig:accrete}), planetesimals have more possibilities to
accrete: with time, as the disk mass decreases, 
the inner disk rapidly moves out of the highly
erosive regime, while erosion still remains important at large orbital
distances. 
With a turbulence strength parameter $\gamma=10^{-4}$, corresponding to
a very weak turbulence (fig.~\ref{fig:accrete_gam1d-4}), the presence
of a highly erosive regime centered around $\sim 300$m planetesimals
is limited only to the outer regions ($\ga 30$ AU), 
and the zone rapidly shrinks
as the circumstellar gaseous disk disappears. 

In order to put these findings into context, we also show the mass of
the disk when Jupiter is believed to have started accreting its
gaseous envelope. These values are calculated by assuming that the
planet growth has been limited mostly by viscous diffusion in the disk,
with the protoplanet capturing between 10\% and 70\% of the mass flux
at its orbital distance in the disk evolution model 
by \citet{Hueso05} \citep[for details, see][]{Guillot06}.
If giant planets have to form, 
at some time 
corresponding to the disk mass interval defined by the hashed areas in
figs.~\ref{fig:accrete_gam1d-2} to \ref{fig:accrete_gam1d-4},
protoplanetary cores must be already large enough
to start accreting the
surrounding hydrogen and helium gas.

We can now define three important disk masses, and their corresponding
disk ages (with the caution that ages are inherently model-dependent
and are provided here for illustrative purposes only, on the basis of
our particular model of $\alpha$-disk evolution with
photoevaporation):
\begin{enumerate}
\item The maximum mass of the disk, following the collapse of the
  molecular cloud. This mass can vary quite significantly from one disk
  formation/evolution model to another. 
  For the particular model shown here, it is
  of the order of $0.25\,\rm M_\odot$, for an age of 0.6\,Myrs. 
\item The disk mass necessary for Jupiter to grow to its present mass
  if it captures 10\% of the mass flux at its orbital distance. For
  realistic disk models, this depends weakly on parameters such as
  $\alpha$ and the disk evaporation rate. In our case, it corresponds
  to $M_{\rm disk}=0.035\,\rm M_\odot$ and an age of 1.95\,Myrs.
\item The disk mass necessary for Jupiter to grow to its present mass
  if it captures 70\% of the mass flux at its orbital distance. For
  our model, $M_{\rm disk}=0.0054\,\rm M_\odot$ (about 5 times the
  mass of Jupiter) and an age of 2.85\,Myrs.
\end{enumerate}

Table~\ref{tab:radii} provides the values of the physical
radii that define the
accretive regime and the highly erosive (disruptive) regime of
figs.~\ref{fig:accrete_gam1d-2} to \ref{fig:accrete_gam1d-4}, namely
$R_{\rm acc}$ and $R_{\rm dis}$.  In our
Solar System, the existence of Jupiter implies that either turbulence
was low, the planet grew from a protoplanetary core formed in the
inner solar system, or a mechanism was able to lead to
the rapid formation of embryos larger than 240 km in radius 
at 5 AU (17 km in
the low-turbulence case, 1080 km in the high-turbulence case) 
by the time when the
disk mass had decreased to $5\times 10^{-3}\,\rm M_\odot$. In the last
case, it appears that a mechanism such as the standard gravitational
instability \citep[e.g.,][]{Safronov69,GW73} would not work
because of the turbulence, but formation of relatively large
protoplanets in eddies or 
vortexes \citep[e.g.,][]{Johansen07,Barge95} is a promising possibility.

\section{Conclusion and discussion}

We have investigated the critical physical radii for collisions between
planetesimals to be accretional ($R > R_{\rm acc}$)
or disruptive ($R < R_{\rm dis}$) in turbulent disks,
as functions of turbulent strength 
($\gamma \sim O(\Delta \rho/\rho)$),
disk gas surface density, and orbital radius. 
The results presented here highlight 
the fact that MRI turbulence poses a great
problem for the growth of planetesimals: generally, only those with
sizes larger than a few hundred km are in a clearly
accretive regime for a nominal value of $\gamma \sim 10^{-3}$. 
The others generally collide with velocities greater than their own surface
escape velocities. 
For some of them, more severely in the
kilometer-size regime, collisions are likely to be disruptive. 
The problem is greater when the disk is still massive and at
large orbital distances.
Also, if turbulence is stronger than $\gamma \sim 10^{-2}$,
planetesimal accretion becomes extremely difficult.

However, the rate of occurrence of
extrasolar giant planets around solar-type stars is inferred to be as
large as $\sim 20$\% \citep{Cumming08}, and depends steeply on
the metallicity of the host star \citep{Fischer05,santos04}. This
strongly suggests that the majority of extrasolar giant planets were
formed by core accretion followed by gas accretion onto the cores
\citep{IL05}.  
Thus, planetesimals should commonly grow to planetary masses
before the disappearance of gas in protoplanetary disks.

The possibilities to overcome the barrier are in
principle as follows (their likelihood is commented below):
\begin{enumerate}
\item Large $M$: 
Large planetesimals with sizes of 100 to 1000\,km are
formed directly in turbulent environment by a mechanism
other than collisional coagulation, jumping over the erosive
regime for physical radii.

\item Small $\Sigma_g$: Planetesimals start their accretion to
  planet-size only after the disk surface density of gas has declined
  to sufficiently small values.

\item Small $\gamma$: Planetesimals form in MRI-inactive regions
  (``dead zones'') of protoplanetary disks.
\end{enumerate}

Concerning point 1, 
the first-born planetesimals with sizes larger than $R_{\rm acc}$ may be formed
rapidly by an efficient capture of $\sim$meter-size boulders in
vortexes \citep{Johansen07}.  Such large planetesimals may be
consistent with the size distribution of asteroids (Morbidelli et
al. 2008).
Even if the first-born planetesimals are not as large, a small
fraction of them could continue to grow larger than 
$R_{\rm acc}$ by accreting smaller bodies, because 
accretion is not completely cut off as soon as
$v_{\rm disp} \ga v_{\rm esc}$
(there is always a small possibility for accretion) and the large
planetesimals would not be disrupted by smaller ones.  This
possibility, however, must be examined by a more detailed growth model
taking into account the effect of fragmentation and the size distribution of
planetesimals, which we neglected in this paper.

Concerning point 2, 
we have shown that planetesimals are most fragile at early times, in
massive disks, and at large orbital distances. We therefore suggest
that the growth towards planet sizes may be delayed due to MRI
turbulence, and then proceed from inside out: planetesimals should
start accretion first close to the star, then progressively at larger
orbital distances, as the gas surface density declines.  The
possibility to delay planet formation while keeping non-migrating
km-size planetesimals is noteworthy because it would help planetary
systems resisting to type-I migration: they would grow in a gas disk
that is less dense, and for which migration timescales may be
considerably increased.  \citet{IL08}, \citet{Alibert05}, and
\citet{Kominami06} showed that type-I migration must be 
lowered by one to two orders of magnitude from
the linear calculation \citep{Tanaka02} to provide an explanation
for the existence of a population of giant planets in agreement with
observations.  This ``late formation'' scenario is 
consistent with the noble gas enrichment in Jupiter
\citep{Guillot06}. 
However, in order to form gas giants,
core accretion and gas
accretion onto the cores must proceed fast enough to capture
Jupiter-mass amount of gas from the decaying gas disk.
Once the size of the largest planetesimals exceeds $\sim 1000$km,
their eccentricities are damped by tidal drag and dynamical friction 
from small bodies.  
Most of the other small bodies
may be ground into sizes smaller than 1\,km and their eccentricities
could be kept very small by gas drag and collision damping.  This
could facilitate the runaway accretion of cores to become large enough
($\ga 10 M_{\oplus}$) for the onset of runaway gas accretion.  This
issue also has to be addressed by a detailed planetesimal growth model
taking into account a size distribution.
The likelihood of relatively rapid gas accretion without long ``phase 2''
is discussed by \citet{Shiraishi}.
If planetesimal sizes are relatively small, 
gas drag damping opens up a gap in the planetesimal disk 
around the orbit of a core and truncates planetesimal accretion 
onto the core.  The truncation of heating due to planetesimal
bombardment enables the core to efficiently accrete disk gas.

Concerning point 3, 
the MRI inactive region (``dead zones'') may exist in inner disk
regions in which the surface density is large enough to prevent cosmic and
X rays from penetrating the disk \citep{gammie96,sano00}.  The
preservation of a dead zone can also contribute to stall type-I
migration by converting it to type-II migration \citep{Matsumura07}
or by creating a local region with a positive radial gradient of disk
pressure near the ice line \citep{IL08b}.  However, dead zones can be
eliminated by turbulent mixing/overshoot
\citep{Varniere06,Turner07,Ilgner08}, a self-sustaining mechanism
\citep{Inutsuka05}, and dust growth \citep{sano00}.  The last effect
comes form the fact that small dust grains are the most efficient
agents for charge recombination.  
According to grain growth, the ionization of the disk and 
its coupling with the magnetic field become stronger to activate
MRI turbulence.
We remark that if MRI turbulence is
activated, collisions are disruptive and they re-produces small grains 
to decrease the ionization degree.
This self-regulation process might maintain a marginally dead state
and keep producing small dust grains.  This might be related with
relative chronological age difference ($\sim 2$Myr) between chondrules
and CAIs \citep[e.g.,][]{Kita05}.  Whether dead zones exist or not is
one of the biggest issues in evolution of protoplanetary disks and
planet formation.  A more detailed analysis of planetesimal accretion in
turbulent disks could impose a constraint on this issue.
 
At large orbital distances (10's of AU), the existence of a highly
erosive regime that lasts until late in the evolution of the
protoplanetary disk is an important feature of this scenario.
It shows that the entire mass of solids is highly reprocessed by
collisions, in qualitative agreement with the paucity of presolar
grains (intact remnants from the molecular cloud core) found in
meteorites.
It also prevents the growth of large
planetesimals and helps to maintain a large population of small grains
in the disks.  This is in qualitative agreement with observations that do
not indicate a significant depletion of micron-sized grains with time,
contrary to what would be predicted in the absence of turbulence 
\citep{Dullemond05,Tanaka05}.

In conclusion, the existence of MRI turbulence may be a threat to
planetesimal accretion.  Given the uncertainties related to these explanations,
we cannot provide a definitive scenario for the formation of
protoplanetary cores. However, it offers several promising hints to
explain important features of planet formation as constrained by
today's observations of protoplanetary disks, exoplanets and
meteoritic samples in the Solar System.

\acknowledgments

This research was supported by the Sakura program
between Japan and France, and by the CNRS
interdisciplinary program {\it ``Origine des plan\`etes et de la
  Vie''} through a grant to T.G. and A.M.

\section*{Appendix}

Here we derive the $r$-dependence in eq.~(\ref{eq:e_random}).
If the equation of motion is scaled by a reference radius $r_1$
and $T_{\rm K1}$ where $T_{\rm K1}$ is a Keplerian period at $r_1$,
the only remaining non-dimensional parameter in the equations is 
\begin{equation}
\gamma \Gamma(r_1) = \gamma \frac{64 \Sigma_g r^2}{\pi M_{\odot}}|_{r_1}
\end{equation} 
(see eqs.~[4], [5], [6] in \citet{ogi07}).
Consider the equation of motion scaled by $r_1$ and $T_{\rm K1}$
and that scaled by $r_2$ and $T_{\rm K2}$.  
If $\gamma$ is the same and $\Gamma(r_1)=\Gamma(r_2)$, 
these two scaled-equations of motion are identical and
evolution of eccentricity, which is a non-dimensional quantity, 
must be identical in terms of the scaled time.
Note that the magnitude of excited $e$ should 
be proportional to $\gamma \Gamma$.     
Since 
\begin{equation}
\Gamma(r) = \frac{\Sigma_g(r)}{\Sigma_g(r_1)} 
            \left(\frac{r}{r_1}\right)^2 \Gamma(r_1),
\label{eq:Gamma_r}
\end{equation} 
and Ogihara et al.'s eq.~(34) derived for
$r = 1$AU is proportional to $\gamma \Gamma(1{\rm AU})$,
the formula for arbitrary $r$ is given by
replacing a year by $T_{\rm K}(r)$ and 
$\gamma$ by $\gamma (\Gamma(r)/\Gamma(1{\rm AU}))$ in their
equation.  As a result,
\begin{equation}
e \sim 0.1 \gamma 
  \left(\frac{\Sigma_g}{\Sigma_{g,1}}\right)
  \left(\frac{r}{1{\rm AU}}\right)^{2}
  \left(\frac{t}{T_{\rm K}}\right)^{1/2},
\end{equation}
where $\Sigma_{g,1}$ is $\Sigma_{g}$ at 1AU with $f_g = 1$
(eq.~[\ref{eq:Sigma_g}]) and the numerical factor was
corrected as explained in the footnote in \S 3.1.
Assuming the simple power-law model defined by eq.~(7),
\begin{equation}
e \sim 0.1 f_g \gamma 
  \left(\frac{r}{1{\rm AU}}\right)^{-1/4}
  \left(\frac{t}{1{\rm year}}\right)^{1/2}.
\end{equation}

\clearpage
\begin{table}
\caption{
Limits of the accretive regime ($R > R_{\rm acc}$) 
and the highly erosive (disruptive) 
regime $R < R_{\rm dis}$, as a function of orbital distance and disk mass.}
\begin{tabular}{r|cc|cc|cc} \hline
 
  & \multicolumn{2}{c|}{1 AU}  &  \multicolumn{2}{c|}{5 AU}  & \multicolumn{2}{c}{30 AU}  \\
\cline{2-3} \cline{4-5} \cline{6-7}
$M_{\rm disk}$ & $R_{\rm acc}$ & $R_{\rm dis}$  & $R_{\rm acc}$ & $R_{\rm dis}$  & $R_{\rm acc}$  & $R_{\rm dis}$ \\
$\rm [M_\odot]$  & [km] & [km] & [km] & [km] & [km] & [km] \\
\hline
\multicolumn{7}{l}{Fiducial case: $\gamma= 10^{-3}$; $\rho_{\rm p}=3\,{\rm g\,cm^{-3}}$; $Q_0=3\times 10^7$; $B=1.0$}\\ \hline
      0.25 &   280. &  $[1.14-0.061]$ &  1280. &  $[82.-0.00047]$ &  1680. &  $[1000.-0.00047]$ \\
     0.035 &   86. &      $--$       &   440. &  $[5.3-0.026]$ &    850.  &  $[200.-0.0024]$ \\
    0.0054 &   46. &      $--$       &   240. &  $[1.05-0.057]$ &    590.  &  $[37.-0.0069]$ \\
\hline
\multicolumn{7}{l}{Low turbulence case: $\gamma= 10^{-4}$; $\rho_{\rm p}=3\,{\rm g\,cm^{-3}}$; $Q_0=3\times 10^7$; $B=1.0$}\\ \hline
     0.25 &    16. &      $--$       &   150. &  $[0.30-0.0099]$ &   590.  &  $[22.-0.0099]$ \\
    0.035 &    3.9 &      $--$       &   36. &     $--$         &   220.  &  $[1.07-0.053]$ \\
   0.0054 &    1.7 &      $--$       &   17. &     $--$         &   103.  &      $--$ \\
\hline
\multicolumn{7}{l}{High turbulence case: $\gamma= 10^{-2}$; $\rho_{\rm p}=3\,{\rm g\,cm^{-3}}$; $Q_0=3\times 10^7$; $B=1.0$}\\ \hline
     0.25 &   2540. &  $[300.-0.0026]$ & 3920. &  $[2770.-2\times 10^{-5}]$ &  4250. & $[3220.-2\times 10^{-5}]$\\
    0.035 &   870.  &  $[26.-0.0090]$ & 1590. &  $[710.-0.0012]$ &            2180. & $[1460.-0.00012]$ \\
   0.0054 &   510.  &  $[7.4-0.018]$ &  1080. &  $[230.-0.0025]$ &            1580. & $[970.-0.00033]$\\
\hline
\multicolumn{7}{l}{High material resistance: $\gamma= 10^{-3}$; $\rho_{\rm p}=3\,{\rm g\,cm^{-3}}$; $Q_0=10^8$; $B=2.0$}\\ \hline
     0.25 &   280.  &  $[0.26-0.15]$  &  1280. &  $[27.-0.0013]$  &  1680. &   $[700.-0.0013]$ \\
    0.035 &   86.  &     $--$        &   440. &  $[1.6-0.068]$ &    850. &   $[66.-0.0068]$ \\
   0.0054 &   46.  &     $--$        &   240. &  $[0.27-0.15]$ &     590. &   $[12.-0.019]$\\
\hline 
\multicolumn{7}{l}{Low material resistance:$\gamma= 10^{-3}$; $\rho_{\rm p}=3\,{\rm g\,cm^{-3}}$; $Q_0=10^7$; $B=0.3$}\\ \hline
     0.25 &   280.  &  $[11.-0.026]$ &   1280. &  $[550.-0.00018]$ & 1680. &   $[1440.-0.00018]$ \\
    0.035 &   86.  &  $[0.56-0.087]$ &   440. &  $[41.-0.011]$    &  850. &   $[590.-0.00097]$ \\
   0.0054 &   46.  &     $--$        &   240. &  $[9.2-0.024]$  &    590. &   $[240.-0.0028]$
\end{tabular}
\label{tab:radii}
\end{table}

\clearpage

\clearpage

\clearpage

\clearpage

\end{document}